\documentstyle[aps,prl,epsfig]{revtex}
\begin{document}
\draft
\twocolumn[\hsize\textwidth\columnwidth\hsize\csname 
@twocolumnfalse\endcsname

\title{On the universal optical conductivity and
single-particle relaxation in cuprates}
\author{P.~Prelov\v sek$^{1,2}$}
\address{ $^{1}$ Institute for Solid State Physics, University
of Tokyo, Kashiwanoha, Kashiwa-shi, Chiba 277-8581, Japan} 
\address{ $^{2}$ Faculty of Mathematics and Physics, University of
Ljubljana, and J.~Stefan Institute, 1000 Ljubljana, Slovenia }

\date{\today}
\maketitle
\begin{abstract}\widetext

  The origin of the universal optical conductivity form $
  \sigma(\omega) \propto (1-{\rm e}^{-\omega/T})/\omega $ established
  in numerical model studies of doped antiferromagnets and consistent
  with experiments on cuprates near optimal doping is discussed. It is
  shown that such a behaviour appears quite generally when the
  single-particle excitations are overdamped and their relaxation
  follows the marginal Fermi liquid concept. Relation to recent ARPES
  experiments is also discussed.
\end{abstract}
]
%\begin{multicols}{2}
\narrowtext

Since the discovery of superconducting cuprates the anomalous linear
resistivity law, $\rho(T) \propto T$, has been the strong indication
that cuprates in the normal state are strange metals even near the
optimum doping not following the normal Fermi liquid
behaviour. Analogous message arises from the analysis of the optical
conductivity \cite{tann} which does not follow the standard Drude form,
\begin{equation}
\sigma(\omega) = \frac {i \omega_p^2}{\omega + i/\tau},
\label{eq1}
\end{equation}
with a constant relaxation rate $1/\tau$ and plasma frequency
$\omega_p$. Experiments can be however well described within the
marginal Fermi liquid (MFL) concept \cite{varm} where the generalized
frequency-dependent rate is introduced, i.e. $\tau^{-1}(\omega,T)=
\tilde \lambda( |\omega| + \eta T)$. These results suggest
that also spectral functions as e.g. measured by the angle-resolved
photoemission spectroscopy (ARPES) have to be anomalous, i.e. the
quasiparticle (QP) relaxation has to follow the MFL dependence. Only
recently the high resolution ARPES experiments on
Bi$_2$Sr$_2$CaCu$_2$O$_{8+\delta}$ (BSCCO) \cite{valla,kami,bogd} seem
to be in position to confirm beyond doubt this behaviour, obeyed in the
optimum-doped materials surprisingly even at $T \alt T_c$ for QP along
the nodal direction in the Brillouin zone.

Recently several static and dynamic response functions at finite $T$
have been studied numerically \cite{jprev} within the prototype
$t$-$J$ model, expected to represent well normal-state properties of
cuprates. It was established that the MFL concept applies to several
dynamic quantities in a broad range of intermediate hole doping
$0.1<c_h<0.3$, in particular to the dynamical conductivity
$\sigma(\omega)$ \cite{jpuni} and to the QP relaxation rate as obtained
from the analysis of spectral functions $A({\bf k},\omega)$
\cite{jpspec}. Moreover, $\sigma(\omega)$ has been been found close to
a parameter-free form \cite{jpuni} (we use $k_B=\hbar =1$),
\begin{equation}
\sigma(\omega) =  C_0\frac{1 - e^{-\omega/T}}{\omega},
\label{eq2}
\end{equation}
in a remarkably broad frequency regime $0<\omega < \omega^* \sim 2t$,
while $C_0$ being essentially $T$-independent for $T<J$.  The
resulting $\sigma(\omega<\omega^*)$ is clearly universal, i.e. the
conductivity is governed by $T$ only. Evidently, Eq.(\ref{eq2})
reproduces the linear resistivity law $\rho=T/C_0$. The form
(\ref{eq2}) is also consistent with the MFL scenario for
$\tau^{-1}(\omega,T)$, however in a very restrictive way since both
MFL parameters are essentially fixed. A reasonable overall fit can be
thus achieved by $\tilde \lambda \sim 0.6$ and $\eta \sim 2.7$, while
for more discussion on MFL parameters emerging from Eq.(\ref{eq2}) we
refer to Ref.\cite{prel}.  When optical experiments on
$\sigma(\omega)$ in cuprates are analysed within the MFL framework
quite close values for $\tilde \lambda, \eta$ are in fact reported
\cite{schl,elaz,puch}. Such a behaviour seems not to be restricted to a
particular model since it has been found also in the analysis of
ladder systems \cite{tsun}, and the prerequisities for its validity
has been partly discussed already in Refs.\cite{jpuni,imad,jprev}.

Our aim is to discuss the relation of $\sigma(\omega)$ and the
associated relaxation rate $1/\tau$ with the single-particle damping
$\Gamma$. In the usual case of a weak scattering both are simply
related, i.e. $1/\tau \sim 2 \Gamma$. Recent ARPES
measurements \cite{valla,kami,bogd} in fact are not consistent with a
weak scattering but rather with overdamped QP excitations, which needs
a reconsideration of usual arguments.

The dynamical conductivity $\sigma(\omega)$ is within the linear response
theory related to the current-current correlation spectral function
$C(\omega)$, which in general replaces the constant $C_0$ in Eq.(\ref{eq2}), 
\begin{equation}
C(\omega) = \text{Re} \int_0^{\infty} dt~ e^{i \omega
t }\langle j(t) j \rangle , \label{eq3}
\end{equation}
where $j$ is the electric current density.  In the following we try to
establish the conditions for the universal behaviour $C(\omega) \sim
C_0$.  The simplest approach is to decouple $C(\omega)$ in terms of
single-particle spectral functions $A({\bf k}, \omega)$,
\begin{eqnarray}
C(\omega)&=& \frac{2\pi e_0^2}{N} \sum_{\bf k} (v_{\bf k}^\alpha)^2 \int
{d\omega^\prime} f(-\omega^\prime) f(\omega^\prime-\omega) 
\nonumber \\ 
&&A({\bf k},\omega^\prime) A({\bf k},\omega^\prime-\omega),
\label{eq4}
\end{eqnarray}
where $f$ is the Fermi function and $v_{\bf k}^\alpha$ unrenormalized
band velocities. Note that Eq.(\ref{eq4}) should become exact for the
local Fermi liquid as realized in the limit of infinite dimensions
\cite{geor}. Relevant for cuprates Eq.(\ref{eq4}) also represents a
convolution of the $A_-$ and $A_+$ spectral functions so the resulting
$C(\omega)$ should scale with the hole doping $c_h$, as expected for
doped Mott insulators.

$A({\bf k}, \omega)$ for QP close to the Fermi
energy ($\omega=0$) are conveniently represented as
\begin{equation}
A({\bf k}, \omega)=\frac{1}{\pi}\frac{Z_{\bf k}\Gamma_{\bf k}}
{(\omega-\epsilon_{\bf k})^2+\Gamma^2_{\bf k}},
\label{eq5}
\end{equation}
where QP parameters $Z_{\bf k},\Gamma_{\bf k},\epsilon_{\bf k}$ in
general dependent on $\omega$ and $T$. In order to reproduce the MFL
form of $\sigma(\omega)$ one has to assume the MFL form for the QP
damping, i.e. $\Gamma = \gamma (|\omega| + \xi T)$, but as well
neglect the ${\bf k}$ dependence of $\Gamma$ and $Z$. We however allow
for the asymmetry of damping of electron-like ($\omega>0$) and
hole-like ($\omega<0$) QP excitations as has been clearly established
in the numerical analysis \cite{jpspec}. We thus consider also cases
where $\gamma_+ \neq \gamma_-$ and $\xi_+ \neq \xi_-$ where $\pm $
refer to $\omega>0$ and $\omega<0$ regimes, respectively. We also neglect
the $Z(\omega)$ dependence within the MFL framework.

Below the cutoff frequency $\omega<\omega^*$ the behaviour near the
Fermi surface should be dominant. Assuming slowly varying
density of states ${\cal N}(\epsilon)$ one can derive \cite{varm}
\begin{equation}
C(\omega)=\bar C \int d\omega' f(-\omega')f(\omega'-\omega)
\frac {\bar \Gamma(\omega,\omega')}
{\omega^2+ \bar \Gamma(\omega,\omega')^2}, \label{eq6}
\end{equation}
where $\bar \Gamma(\omega,\omega')=\Gamma(\omega')
+\Gamma(\omega'-\omega)$. depending only on $\omega/T$ and MFL
parameters $\gamma,\xi$, where $\xi\sim \pi$ is usually used.

It is evident from Eqs.(\ref{eq6}) that for $\gamma \ll 1$
one recovers $C(\omega)$ strongly peaked at $\omega = 0$ and
consequently $C(\omega) \propto \sigma(\omega) \propto \bar
\Gamma/(\omega^2+ \bar \Gamma^2)$ with $\bar \Gamma=
2\Gamma(\omega/2,T)$. This is just a generalized Drude form with
$1/\tau(\omega)=2\Gamma(\omega/2)$, invoked in connection with
the MFL concept \cite{varm}.

For the regime of overdamped QP with $\gamma \sim 1$ or more
appropriate $\gamma \xi \sim 1$ we present in Fig.~1 $C(\omega)$ for
several $\gamma$ fixing $\xi=\pi$. While for $\gamma \alt 0.2$ still a
pronounced peak shows up at $\omega \sim 0$, $C(\omega)$ becomes for
larger $\gamma$ nearly constant or very slowly varying in a wide range
of $\omega/T$. For $\gamma>0.3$ one finds $C(0)<C(\omega \to \infty)$,
approaching for $\gamma \gg 1$ the ratio $C(0)/C(\infty)=1/\xi$.
\begin{figure}
\centering 
%\epsfxsize=7 cm
%$$\epsffile{figmfn1.eps}$$
\epsfig{file=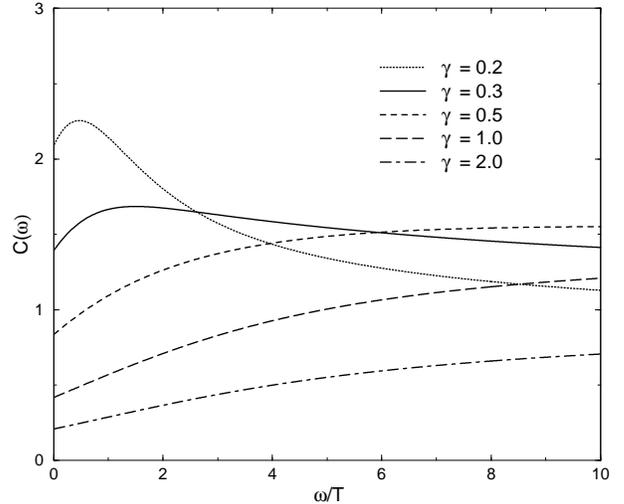,height=8cm,angle=-90}
\caption{ Current-current correlation spectra $C(\omega)$
vs. $\omega/T$ for various $\gamma$ at fixed $\xi= \pi$.}
\label{fig1}
\end{figure}

Quite similar is the behaviour in the case of an asymmetric QP
damping. In Fig.~2 we present some results for fixed $\gamma_-=0.7$
and $\xi=\pi$, while varying $\gamma_+$. From Eq.(\ref{eq6}) it is
evident that essentially $\gamma_+ +\gamma_-$ matters, so quite
constant $C(\omega)$ appears already for $\gamma_+ \agt 0.2$.
\begin{figure}
\centering
%\epsfxsize=7 cm
%$$\epsffile{figmfn2.eps}$$
\epsfig{file=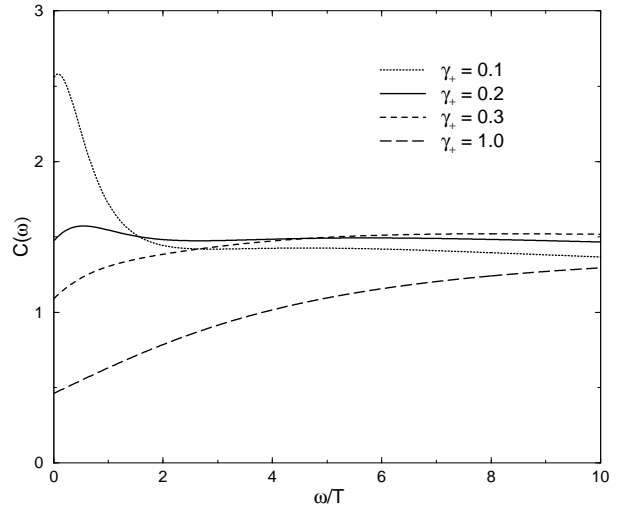,height=8cm,angle=-90}
\caption{$C(\omega)$ vs. $\omega/T$ for various $\gamma_+$ at fixed
$\gamma_-=0.7$. }
\label{fig2}
\end{figure}

The main message of such a simple analysis is that for systems with
overdamped QP excitation of the MFL form Eq.(\ref{eq2}) describes
quite well $\sigma(\omega)$ for a wide range of parameters.  Nearly
constant $C(\omega<\omega^*)$ also means that the current relaxation
rate $1/\tau^*$ is very fast, $1/\tau^* \sim \omega^* \gg 1/\tau$,
i.e. much faster than the conductivity relaxation scale apparent from
Eqs.(\ref{eq1},\ref{eq2}) where $1/\tau \propto T$ is determined
solely by thermodynamics. It should be mentioned that one can fit
results for large $\gamma$ also with Eq.(\ref{eq1}) with large $\tilde
\lambda$ \cite{varm}, but in this regime the form of $\sigma(\omega)$
is quite insensitive to $\tilde \lambda$. 

It is evident that the preceding treatment of $\sigma(\omega)$ is
oversimplified. Vertex corrections seem to be rather unavoidable in
the treatment of strongly correlated electrons in lower dimensions.
Delicate appears the assumption that QP parameters $\Gamma_{\bf
  k},Z_{\bf k}$ are independent of ${\bf k}$ near the Fermi surface.
However, for the validity of Eq.(\ref{eq6}) it is essentially enough
to assume that $\Gamma_{\bf k}(\omega)$ is independent of deviations
$\Delta {\bf k}_{\perp}$ perpendicular to the Fermi surface. In fact
this is just what is observed in recent ARPES studies of BSCCO
\cite{valla,kami}.  A smooth depedence on $\Delta {\bf k}_\parallel$
which on the other hand appears to be quite pronounced
\cite{valla,bogd} could be presumably acommodated not spoiling general
conclusions.  One should however realize the importance of the upper
cutoff scale $\epsilon^*$ for the validity of the MFL QP behaviour in
Eqs.(\ref{eq4},\ref{eq5}). In the regime of large $\gamma\xi \gg 1$,
which appears to the case in cuprates, even at moderately low $T$
$C(\omega)$ could be influenced by cutoff $\epsilon^*$.

It seems that more complete calculations of $C(\omega)$, as e.g.
performed numerically \cite{jprev,tsun} lead to even flatter
$C(\omega<\omega^*)$, than found in Figs.~1,2 e.g. for $\gamma \gg 1$.
In fact the condition for $C(\omega<\omega^*) \sim C_0$ is that the
decay of $C(t)$ is fast and monotonous. Within the planar $t$-$J$
model at the intermediate doping it was found that $\omega^* \sim 2t$,
allowing for an effective mean free path $l^*$ of only few cells. Such
a short $l^*$ can be plausibly explained by assuming that charge
carriers - holes entirely loose the phase coherence in collisions with
each other due to randomizing effect of an incoherent spin background.
Note again that short correlation length (even at $T \alt T_c$)
appears also from the analysis of ARPES spectral functions $A({\bf
  k},\omega)$ varying $\Delta {\bf k}_\parallel$ along the Fermi
surface \cite{valla,bogd}.

Let us finally discuss in more detail the relevance of the above
analysis to the situation in cuprates, as well as the relation to
numerical model calculations performed for the $t$-$J$ model at
$T>0$. Recent analysis of ARPES results in BSCCO gives for hole-like
excitations in the particular (nodal) direction $(0,0) - (\pi,\pi)$
the MFL form with $\Gamma \sim 0.75 \omega$ for $\omega>T$ and $\Gamma
\sim 2.5 T$ for $\omega<T$ \cite{valla}. Similar is the analysis of
spectra in Refs.\cite{kami,bogd}. This means definitely an overdamped
character of hole excitations, since the full width at half maximum
(FWHM) $\Delta \sim 2\Gamma(\epsilon)>\epsilon$ is always larger than
the QP (binding) energy $\epsilon$. Also one should note that the QP
damping appears even larger along other parts of the FS. Quite similar
QP damping was established within the $t$-$J$ model
\cite{jpspec,jprev} where at intermediate doping the hole-part self
energy was found to be of the MFL form, i.e. Im$\Sigma \sim
-\gamma(\omega+\xi T)$ with $\gamma \sim 1.4$ and $\xi \sim 3.5$. In
making the comparison one should take into account that $\Gamma = Z
|{\rm Im}\Sigma|$. Since at the peak position we find $Z \sim 0.5$
experimental and model values appear reasonably close. If one takes any
of these data they do not satisfy the weak-scattering relation $\tilde
\lambda \sim 2\Gamma $, since optical experiments give $\tilde \lambda
\sim 0.5-0.7 < 2 \gamma$ \cite{schl,elaz,puch}.

In conclusion, we have investigated the origin of the universal
conductivity Eq.(\ref{eq2}) which represents a novel diffusion law,
being a counterpart of the usual Drude form for weak scattering, but
here determined bty $T$ only. From the analysis it follows
that such a behaviour generally appears when the QP are overdamped and
their relaxation also follows the MFL behaviour. The calculated
spectra $C(\omega)$ represent a correction to the universal form but
show rather insignificant variation under the latter conditions.
Morerover one can expect its validity even in a more elaborate
treatment.  While the case of cuprates at optimum doping is striking
due to a large range of validity (both in frequency and temperature)
of the novel dynamics the situation is valid in more restricted range
also outside the optimum doping regime (where additional low energy
scales appear) and possibly also in other strongly correlated systems.

The author wishes to thank J.C. Campuzano, P.D. Johnson and T.M. Rice
for helpful discussions, and acknowledges the support of the 
Japan Society for the Promotion of Science, and also of 
the Institute for Theoretical Physics, ETH Z\"urich, and the Institute
for Materials Research, Tohoku University Sendai, where the part
of this work has been done.

%\end{multicols}
\end{document}